\documentclass[structabstract]{aa}

\usepackage{graphicx}
\usepackage{natbib}
\usepackage{amsmath}
\usepackage{amssymb}
\usepackage{eurosym}
\usepackage{latexsym}

\begin{document}

   \title{Observations of a solar flare and filament eruption in Lyman $\alpha$ and X-rays }

   \author{F.~Rubio~da~Costa
          \inst{1,2}
          \and
          L.~Fletcher
          \inst{1}
          \and 
          N.~Labrosse
          \inst{1}
          \and 
          F.~Zuccarello
          \inst{2}}

   \institute{Department of Physics and Astronomy,
              University of Glasgow, 
              Glasgow, G12 8QQ, 
              U. K.\\ 
              \email{frdc@oact.inaf.it}
         \and
	      Department of Physics and Astronomy,
	      University of Catania,
	      Via S. Sofia 78, 95123, 
              Catania, Italy.\\}

   \date{Received; accepted }

\newcommand{\lya}{L$\alpha$}
\newcommand{\ha}{H$\alpha$} 
 
  \abstract
   {\lya\ is a strong chromospheric emission line, which has been relatively rarely observed in flares. The Transition Region and 
Coronal Explorer (TRACE) has a broad ``Lyman~$\alpha$'' channel centered at 1216~\AA\, used primarily at the beginning of the mission. A small number of flares were observed in this channel.}
   {We aim to characterise the appearance and behaviour of a flare and filament ejection which occurred on 8th September 
1999 and was observed by TRACE in \begin{bf}L$\alpha$\end{bf}, as well as by the  \emph{Yohkoh} Soft and Hard X-ray telescopes. We explore the flare energetics and its spatial and temporal evolution. We have in mind the fact that the \lya\ line is a target for the Extreme Ultraviolet Imaging telescope (EUI) which has been selected for the Solar Orbiter mission, as well as the LYOT telescope on the proposed SMESE mission.}
{We use imaging data from the TRACE 1216~\AA, 1600~\AA\ and 171~\AA\ channels, and the \emph{Yohkoh} hard and soft X-ray telescopes. A correction is applied to the TRACE data to obtain a better estimate of the pure \lya\ signature. The \lya\ power is obtained from a knowledge of the TRACE response function, and the flare electron energy budget is estimated by interpreting \emph{Yohkoh}/HXT emission in the context of the collisional thick target model. }
   {We find that the \lya\ flare is characterised by strong, compact footpoints (smaller than the UV ribbons) which correlate well with HXR footpoints. The \lya\ power radiated by the flare footpoints can be estimated, and is found to be on the order of $10^{26}$~erg s$^{-1}$ at the peak. This is less than 10\% of the power inferred for the electrons which generate the co-spatial HXR emission, and can thus readily be provided by them. The early stages of the filament eruption that accompany the flare are also visible, and show a diffuse, roughly circular spreading sheet-like morphology, with embedded denser blobs.}
    {On the basis of this observation, we conclude that flare and filament observations  in the \lya\ line with the planned EUI and LYOT telescopes will provide valuable insight into solar flare evolution and energetics, especially when accompanied by HXR imaging and spectroscopy.}

   \keywords{Sun: activity -- Sun: filaments -- Sun: X-rays, gamma rays -- Sun: coronal mass ejections (CMEs)}

   \maketitle

\section{Introduction}
Early observations of solar flares, prior to the advent of space-based astronomy, recognised them as chromospheric 
phenomena. Since the 1990s, coronal observations show dramatic evidence of heating particle, acceleration and magnetic 
field reorganisation, but it is clear that most of the flare energy (as distinct from the kinetic energy of any 
possible coronal mass ejection) is ultimately radiated away as chromospheric emission, predominantly hydrogen Balmer~
$\alpha$ (more generally known as~H$\alpha$), hydrogen Lyman~$\alpha$  (\lya) and the UV to infrared continuum.  Observations at extreme UV and X-ray wavelengths are essential and show exciting and dynamic  coronal evolution, but to form a complete picture of the flare phenomenon we must integrate them with the perhaps less glamorous L$\alpha$, \ha\ and broadband observations of the chromosphere.

One of the overriding questions in solar flare physics remains the conversion of magnetic energy stored in the corona to 
non-thermal electrons and ions, heat and mass motion. In the context of the chromosphere, the standard model for a flare 
(particularly the `collisional thick target' electron beam heating model) proposes that the chromospheric flare heating 
and enhanced radiation arises as a result of energy deposition and ionisation by a beam of electrons accelerated in the 
corona. In terms of the timing relationship and the energetics the two are compatible, although explaining the broadband 
white light emission proves problematic since it is not clear how an electron beam can penetrate to the depths required 
for the enhanced  H$^-$. However, the electron beam model has thus far proved reasonably successful in the context of 
both \lya\ and \ha\ observations \citep[e.g.][]{2007A&A...461..303R} and we use this as a working hypothesis in interpreting the observations presented here.

Hydrogen lines (\lya\ and H$\alpha$) and continua are a significant source of chromospheric cooling during flares, together with H$^-$ and heavy ion lines and continua \citep{1980sfsl.work..231C}, accounting for a few percent of the overall chromospheric cooling.  A few observations of flares in \lya\ exist, but these are mostly in spectroscopic mode, with no spatial information.  \citet{1980sofl.symp..451C}, studied a flare of soft X-ray importance M1 and \ha\ class 1N, which occurred on 5 September 1973, and was observed by two spectrometers on Skylab, and reported a \lya\ radiative power output of $2.2 \times 10^{25}~{\rm erg~s}^{-1}$. This accounted for on the order of 5\% of the total radiative output at the maximum of the flare.  Using data from the OSO-8 satellite,  \cite{1984SoPh...90...63L} measured a maximum \lya\ intensity of $2.1 \times 10^6$~erg cm$^{-2}$ s$^{-1}$ sr$^{-1}$ in a flare of GOES class M1.0 on 15 April 1978. \cite{1978SoPh...60..341M} observed a number of events with the  Harvard College Spectroheliometer on Skylab, which had a slit width of 5", and reported \lya\  line flux integrated over the line of $\sim 6 \times 10^6$~erg s$^{-1}$ cm$^{-2}$ sr$^{-1}$, averaged over bright \lya\ patches in a 2B flare on September 7 1973.

Following these early reports, \lya\ observations lay relatively dormant. A fortuitous observation by 
\cite{1996ApJ...468..418B} from the Solar Stellar Irradiance Comparison Experiment
\citep[SOLSTICE,][]{1993JGR....9810667R, 1993JGR....9810679W} showed strong enhancements in UV lines, apart from \lya, the irradiance of which (i.e. the total flux density from the entire solar disk) only increased by 6\% above its non-flaring value, compared to resonance lines of \ion{S}{iv} and \ion{C}{iv} which had irradiance increases of a factor 12. However, the \lya\ line was observed at the end of the impulsive phase, whereas the \ion{S}{iv} and \ion{C}{iv} lines were observed near the impulsive peak. The \lya\ radiance (specific intensity of a resolved feature) was estimated to be some 75 times stronger than its 
pre-flare average disk value, and despite the small increase \lya\ was found to be more energetically important than the 
other resonance lines.  The strongest \lya\ enhancement was in the line wings, with the blue wing somewhat more enhanced. The changes in irradiance were found to lie between the values predicted by the F1 and F2 flare models of 
\cite{1980ApJ...242..336M}.

More recently, `Sun as a star' measurements were made with the VUSS instrument onboard the CORONAS-F satellite, and the total \lya\  irradiance increase estimated to be about 10\% \citep{2006SoSyR..40..111N} in an X-class flare. Comparing with data from the SONG instrument on the same spacecraft, the \lya\ peaks are shown to be `synchronous' with hard X-ray emission at tens to hundreds of keV  
\citep{2006SoSyR..40..282N}, although the degree of synchronisation is not discussed.
By comparison, `Sun as a star' measurements performed by SOHO/SUMER gave a relative signal increase of 70\% at the head of the hydrogen Lyman continuum (910~\AA) for an X5.3 flare \citep{2004A&A...418..737L}.
  
The Transition Region and Coronal Explorer \citep[TRACE,][]{1999SoPh..190..351H} has four UV channels, one of which includes the \lya\ transition. Though usually called the \lya\ channel, it has a broad spectral response, from just below 
1100~\AA\ to somewhat above 1600~\AA, and thus has a significant contribution from the \ion{C}{iv} lines at 1548~\AA. In the quiet Sun, this can be corrected to yield the ``pure'' \lya\ intensity based on an empirical method developed for the quiet Sun \citep{1999SoPh..190..351H}. A small number of flares observed in the \lya\ passband early in the TRACE mission have been reported in the literature. For example \cite{1999SoPh..187..261S} and \cite{2001SoPh..198..325S} discuss the downflows observed in \lya\ during the late phase of a flare, and  \cite{2001ApJ...554..451F} examine 1216~\AA~ejecta and chromospheric source movements at low cadence, just after the impulsive phase.  \cite{2001ApJ...560L..87W} mention that \lya\ pre-flare brightenings are not co-spatial with subsequent flare hard X-ray emission sites. However, to our knowledge no detailed analysis of TRACE \lya\ channel observations spanning the impulsive phase have been presented. We will explore the information available 
from the TRACE 1216~\AA~ channel alone, and in combination with the TRACE 1600~\AA\ channel, estimating the flare \lya\ intensity for comparison with previous work.

The \lya\ line has been of interest in flares as a possible diagnostic of proton beams. \cite{1976ApJ...208..618O} 
proposed that charge exchange between beam protons and chromospheric hydrogen, and emission from the moving hydrogen atoms that result, would lead to enhanced emission in the wings of the chromospheric lines. More refined calculations \citep[e.g.][]{1985ApJ...295..275C,1995A&A...297..854F,1998A&A...330..351Z} suggested that this should be an observable effect. Early Skylab observations of the \lya\ line profile by \cite{1980SoPh...67..339C} detected a slight red asymmetry in the early phase of a flare, however the profiles were not in agreement with models of \cite{1985ApJ...295..275C}, and the SOLSTICE observations of \cite{1996ApJ...468..418B} showed a stronger \lya\ blue wing. (A related search for 
alpha-particle beams using \ion{He}{ii} \lya\ emission by \cite{2001ApJ...555..435B} also had a null result).
These discrepancies could be explained in terms of various inclinations of the proton beam with respect to the magnetic field and of the line-of-sight \citep{1998A&A...330..351Z}. \cite{1995A&A...297..574H} and \cite{1998A&A...330..351Z} conclude from their theoretical studies that non-thermal charge-exchange \lya\ (and Ly$\beta$) emission due to proton beams could be detected best at the very early stages of the flare when the coronal column mass is small. Later in the flare, the proton energy at the layers emitting the hydrogen spectrum is reduced by collisions.

On this basis, it seems plausible that line shifts and wing enhancements in the \lya\ line profiles observed at later stages are due to bulk velocities of hydrogen atoms in the flaring atmosphere. There are few examples of theoretical studies incorporating both the effects of a non-thermal electron beam and of flows on the shape of the \lya\ line profile. We refer the reader to the radiative hydrodynamic models of \cite{2005ApJ...630..573A}.

This paper aims to characterise the appearance and behaviour of a flare and filament ejection observed by TRACE in L$\alpha$ as well as by the \emph{Yohkoh} Soft and Hard X-ray telescopes.
In Section~\ref{sect:data} we review the observations and describe the data 
processing, and in Section~\ref{sect:overview} we give an overview of the flare event. Energy diagnostics in \lya\ and hard X-ray from the impulsive phase are discussed in Sections ~\ref{sect:uvenergetics} and ~\ref{sect:hxrenergetics} respectively. The flare was accompanied by highly structured  ejecta, and this is presented in Section~\ref{sect:dynamics}. We present our discussion and conclusions in Section~\ref{sect:disc}.

\section{Data analysis}\label{sect:data}
In this section we discuss observations of an M1.4-class flare on 1999 September 08, in active region NOAA 8690, which was observed by the Transition Region and Coronal Explorer \citep[TRACE;][]{1999SoPh..187..229H}  and the \emph{Yohkoh} Soft X-ray Telescope \citep[SXT;][]{1991SoPh..136...37T} and Hard X-ray Telescope \citep[HXT;][]{1991SoPh..136...17K}. The flare onset was 11:58 UT with a peak at 12:13 UT. 

\subsection{TRACE data}
The TRACE telescope primary and secondary mirrors are divided into quadrants, three of which are coated with multilayers for imaging at EUV wavelengths.
The fourth quadrant is coated with aluminium and magnesium fluoride for imaging very broad ultraviolet wavelength ranges 
near 1216, 1550, 1600 and 1700~\AA. Images in all the wavelengths are projected onto a single CCD detector with pixels 
of 0.5'' on a side, corresponding to a 1.0" spatial resolution.
Our flare was observed with an image size of $768 \times 768$ pixels, and  1216 and 1600~\AA\ images were interleaved with a cadence of approximately 5~s per image. At the very beginning and end of the event we also have some 171~\AA\ images.

The original images taken by the TRACE telescope must be corrected for instrumental effects. The standard procedures included in the Solar Software \citep[SSW;][]{1998SoPh..182..497F} IDL routine trace\_prep  provided by the TRACE team are subtraction of the dark pedestal (Analog-Digital-Converter ADC); flat-field correction; correction for exposure time; image normalisation to DN per sec; and alignment and pointing correction between the two UV channels. The TRACE data were also corrected for solar rotation.
In order to compare TRACE images with \emph{Yohkoh} data, we made a TRACE pointing correction, based on co-aligning the strongest TRACE and  \emph{Yohkoh} HXT sources, assumed to be flare footpoints. We estimate that this can be done with an accuracy of 5". The SXT pointing with respect to HXT is accurate within 1''. 

\begin{figure}
\centering
  \includegraphics[width=8.5cm]{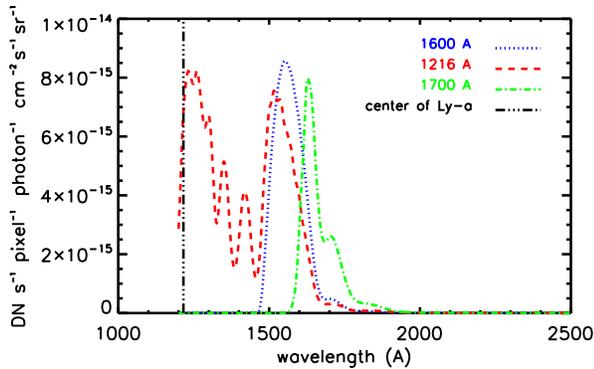}
     \caption{Response of the TRACE 1216~\AA\ and 1600~\AA\ channels. The 1600~\AA\ filter curve has been normalised to the response of 1216~\AA\ filter for comparison of the respective bandpasses.}
     \label{fig:response}
\end{figure}

As shown in Figure~\ref{fig:response}, the 1216~\AA\ channel has a broad spectral response, having approximately the same sensitivity in the $\sim$1500-1600~\AA\ range as it does at $\sim$ 1216~\AA\ \citep[see][]{1999SoPh..187..229H}. This unusual spectral response results from the convolution of a narrowband UV coating on the primary mirror at 1550~\AA\ and a filter near the focal plane centered at 1216~\AA. Much of the contribution to the 1216~\AA\ channel in the quiet Sun is continuum. The 1500-1600~\AA\ range, for which the 1600~\AA\ channel is optimised, includes the strong \ion{C}{iv} pair, as well as lines of \ion{Si}{ii} and continuum. It has essentially no response at wavelengths less than 1400~\AA.

As we can see in Fig. \ref{fig:response}, around 60\% of the response in the 1216~\AA\ channel is from the  \lya\ 1216~\AA\ emission line while the balance is from UV emission near 1550~\AA\  and longer wavelengths. 
\citet{1999SoPh..190..351H} provide a simple method to remove this contamination by using a large number of solar
irradiance data samples from \emph{Solar-Stellar Irradiance Comparison Experiment}  
\citep[SOLSTICE;][]{1993JGR....9810667R}, and by comparing TRACE 1216~\AA~images with \lya\ images from the \emph{Very-high resolution
Advanced ULtraviolet Telescope} \citep[VAULT;][]{2001SoPh..200...63K} sounding rocket flight of May 7, 1998.
Their result is that pure \lya\ intensity can be obtained by a linear combination of TRACE 1216~\AA\ and 1600~\AA\ channels:
\begin{equation}
\centering
I_{L\alpha}=A\times I_{1216}+B\times I_{1600},
\label{eq1}
\end{equation}
where $I_{L\alpha}$ is the intensity of the `pure' (i.e. corrected)  \lya\ emission line, $I_{1216}$ and $I_{1600}$ are the intensities as 
observed with TRACE 1216~\AA\ and 1600~\AA\ channels, respectively, and $A$ and $B$ are the fitting parameters. 
\citet{2006A&A...456..747K} found best-fit coefficients of $A=0.97$ and $B=-0.120$, while \citet{1999SoPh..190..351H} found $B=-0.14$. These coefficients apply to quiet Sun observations, and it is not clear that they can be applied to flare observations, in which the ratio of contributions in each of the TRACE channels may be different from their quiet Sun values, depending on the line and continuum excitations. 
The SOLSTICE flare observations of \cite{1996ApJ...468..418B} showed a proportionally much larger increase in \ion{C}{IV} emission than in \lya. The continuum in both regions increased by about a factor 2. However as noted above, because of the direction of the wavelength scan the \lya\ intensity was measured by Brekke \emph{et al.} during the decay phase, whereas the \ion{C}{IV} intensity was measured in the impulsive phase. In the absence of any further information we will use the \cite{2006A&A...456..747K} prescription.
The effect of the correction applied can be seen by comparing the light curves shown in Fig. \ref{fig1}, and the top panel of Fig. \ref{fig5} with Fig. \ref{fig8}.

\subsection{\emph{Yohkoh} data}
The soft X-ray telescope (SXT) on \emph{Yohkoh}  is a grazing-incidence telescope with a  nominal spatial resolution of about 5'' (2.46" pixels). It makes images in six different filters; we will present images from the Be 119 $\mu$m filter and the Al 12 filter. The \emph{Yohkoh} Hard X-ray telescope (HXT) uses a Fourier-synthesis technique to make images in four hard X-ray (HXR) energy bands i.e., L (14-23 keV), M1 (23-33 keV), M2 (33-53 keV) and H (53-93 keV). The spatial resolution of HXT is described by the collimator response FWHM of about  8''. For our event, \emph{Yohkoh} entered flare mode at about 12:13:36 UT. The flare response provided full-resolution SXT images of a region 2.6 x 2.6 arcmin$^2$ at a cadence of about 2 s. High-cadence HXT data was also taken. Unfortunately, coverage stops at 12:14:30,  shortly after the flare mode onset, due to a data downlink. However there are sufficient data to reconstruct three images.

The \emph{Yohkoh}/SXT data are prepared using the SolarSoft sxt\_prep routine to correct the saturated pixels, subtract the dark current, despike and correct for the vignette loss in the off-axis signal. Exposure normalisation is performed, resulting in images normalised to DN/sec/image pixel. Saturated pixels are flagged. Data are separated by channel and image resolution. The resulting counts per second per image pixel are used to produce both images and light curves for the event. 

\subsection{Co-alignment between TRACE and \emph{Yohkoh} observations}
We co-align TRACE images with SXT to examine the distribution of high-temperature plasma (Fig.~\ref{fig4}). It is well known that TRACE has a pointing drift during its orbit. To compare the high energy data from SXT and HXT on TRACE, the SSW routine TRACE\_MDI\_ALIGN was used, which correlates TRACE white light and the nearest-in-time MDI white light data. It is then assumed that MDI and SXT are well aligned, and that the relative pointing of the two spacecraft does not drift significantly before the flare observation starts. The nearest in time TRACE and MDI white light images occur at 12:01:22~UT and 12:00:00~UT respectively, and the start of the flare observation is 5 minutes later. There will likely be some relative drift in the intervening period but we do not expect this to be significant.

\section{Overview of the event}\label{sect:overview}

\subsection{Time evolution}\label{sect:time}

\begin{figure}
\centering
  \includegraphics[width=9.1cm]{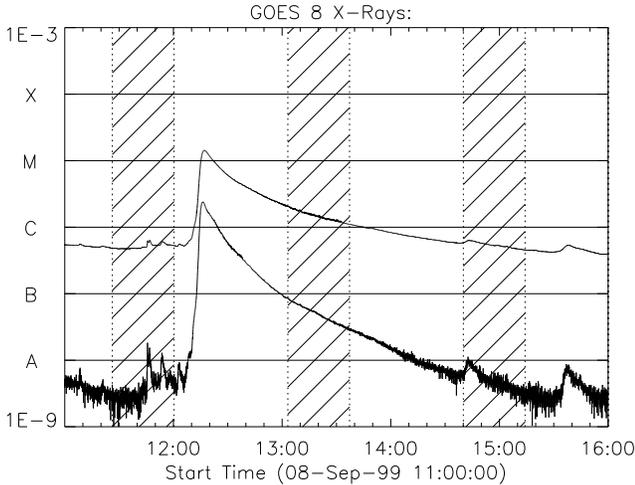}
     \caption{Temporal variation of the 3-second X-ray flux measured with GOES. The upper curve corresponds to the GOES 1.0-8.0~\AA~data, and the lower to the 0.5-4.0~\AA~ data. Hashed regions indicate \emph{Yohkoh} night.}
     \label{fig1b}
\end{figure}

The flare studied here was catalogued as an M1.4 class flare, beginning on 1999 September 08 at
11:58:00 UT. The GOES maximum was detected at 12:13:20 UT and the duration was around 50 minutes. The GOES light curves  for the event are presented in Figure~\ref{fig1b}.  It is possible to differentiate three different phases of the flare:
\begin{itemize}
\item Pre-flare phase from 12:10:00 to 12:13:00~UT: The soft X-ray and UV emission begins to increase. Hard X-ray emission is not yet detected.
\item Impulsive phase from 12:13:00 to around 12:16:00~UT: The soft X-ray flux emission rises more rapidly; the hard X-ray emission increases with several spikes over ten of seconds (See Fig. \ref{fig3} inset for more detail). The UV emission peaks.
\item Post-flare phase from 12:16:00 UT onwards: The soft X-ray flux rises to a peak at 12:16:00, then decays exponentially during several hours. No hard X-ray emission is seen during this time. 
\end{itemize}

\begin{figure}
\centering
  \includegraphics[width=9.1cm,height=8cm]{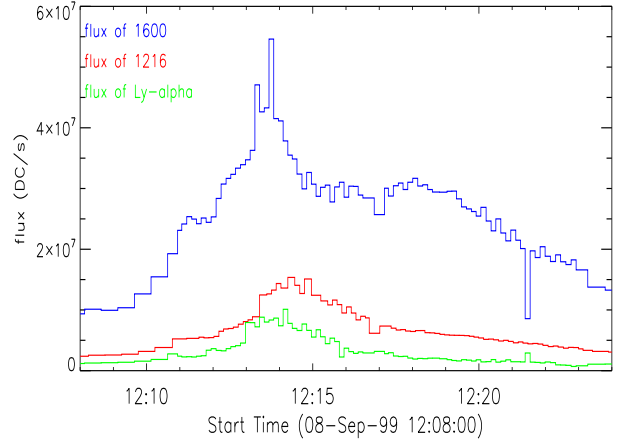}
     \caption{Temporal evolution of the TRACE 1216~\AA, TRACE 1600~\AA\ and the \lya\ corrected flux.}
     \label{fig1}
\end{figure}

\begin{figure*}
\centering
\includegraphics[width=16.5cm]{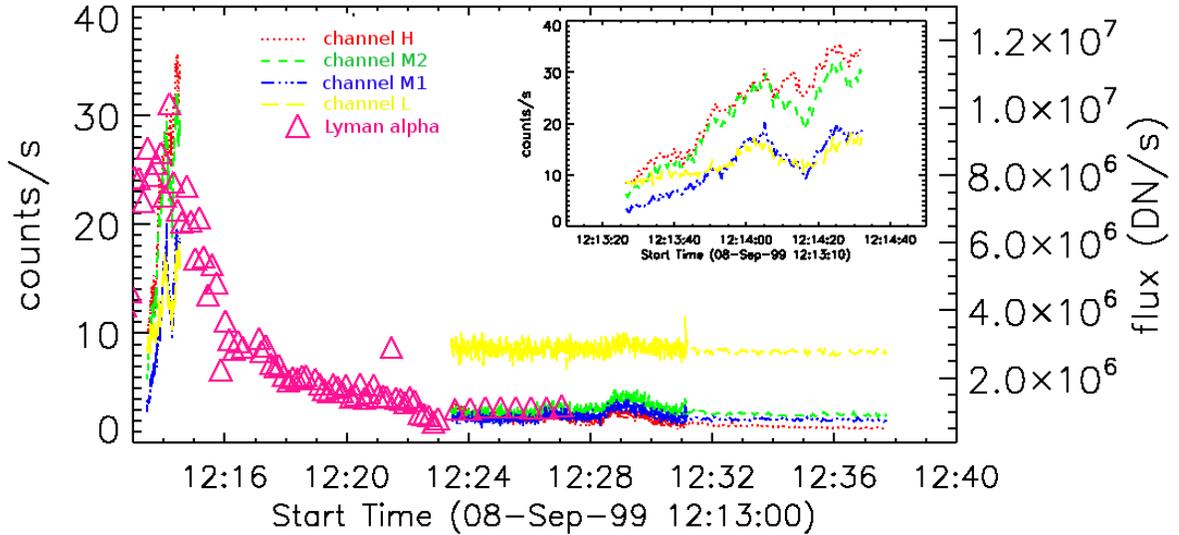}
 \caption{Time profile of corrected TRACE \lya\ flux (see Section 3.3) shown as pink triangles, and Hard X-Ray counts/subcollimator/second taken in the early impulsive phase with \emph{Yohkoh}/HXT. The four different channels represented are 14-23 keV (L), 23-33 keV (M1), 33-53 keV (M2) and 53-93 keV (H). The inset shows the HXT time profiles at the beginning of the flare at higher time resolution}
     \label{fig3}
\end{figure*}

Fig. \ref{fig1} shows the time profiles of the TRACE 1216~\AA\ and 1600~\AA\ channels, and the corrected \lya, in counts per second. These were obtained by thresholding the normalised exposures at a level of 4000 counts per second, thus isolating the brightest flare sources. The footpoint sources were very intense, but saturation was mostly avoided as the TRACE automatic exposure control limited the exposure times to between 0.2 and 0.3~seconds in the 1216~\AA~ channel, and between 0.04 and 0.1~seconds in the 1600~\AA~channel. The 1600~\AA\ flux rises earlier and more rapidly than the 1216~\AA\ flux, and also peaks $\sim$ 1 minute earlier. 
The TRACE 1216~\AA\ light curve decays more rapidly. During the flare observed by \citet{1996ApJ...468..418B} the \lya\ line remained stronger overall, although the \ion{C}{IV} lines increased in intensity by a much larger factor than did the \lya\ line. We can deduce that the \ion{C}{IV} contribution to the 1216~\AA\ channel during the flare is rather small, and the differences between the two light curves  primarily reflect the behaviour of the \lya\ and \ion{C}{IV} lines. There is evidence for  more early phase emission in \ion{C}{IV}, with a somewhat more impulsive behaviour. This might indicate that the atmospheric layers responsible for producing the \ion{C}{IV} emission are heated/excited before the \lya\ emitting layers.
The corrected \lya\ also rises earlier than 1600~\AA\ or 1216~\AA. This \lya\ time information must be treated with caution because we time-interpolated intensities in the two TRACE filters, and because the correction we applied might not correctly estimate the \ion{C}{IV} contribution, and bias the corrected \lya\ light curve.

The HXT light curves from the early impulsive phase and in the decay phase are shown in Figure ~\ref{fig3}.  Only the earliest part of the impulsive phase was observed before the \emph{Yohkoh} data downlink, followed by spacecraft night. The flare was in its decay phase when observations started again. The \emph{Yohkoh}/SXT light curve (not shown) demonstrates the usual steep rise in soft X-rays at the beginning of the flare, and then the slow decay as the spacecraft comes out of eclipse. 

\subsection{Morphology of the event}

\begin{figure}
\centering
\hbox{
 \includegraphics[width=4.8cm]{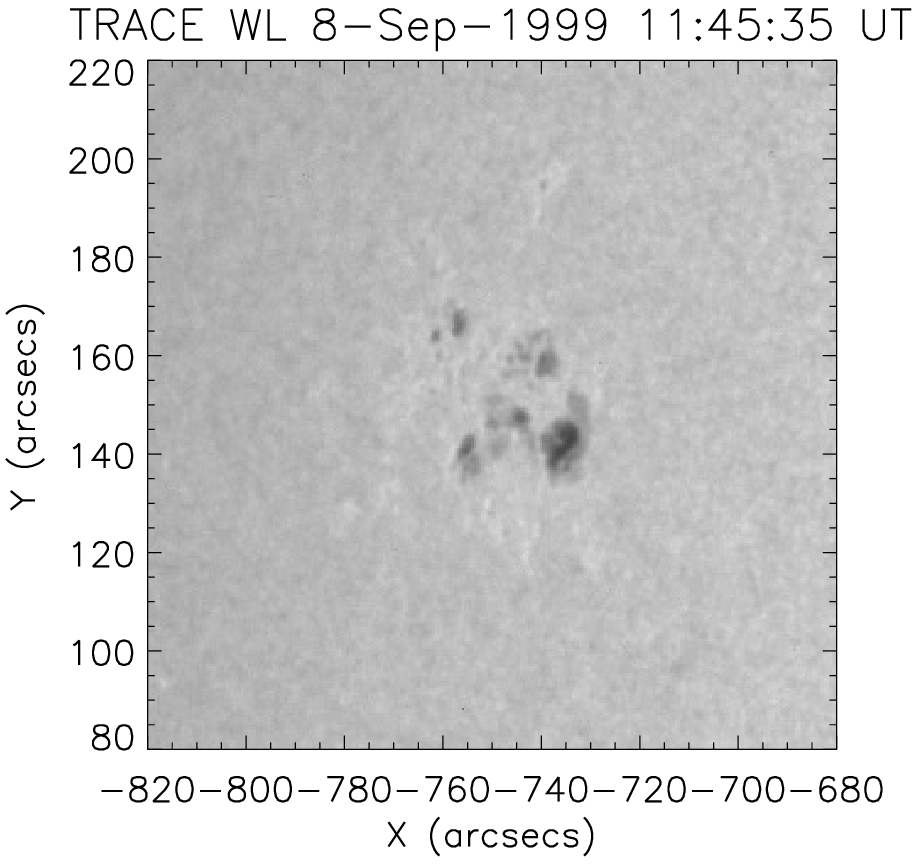}
 \hspace{-0.5cm}
 \includegraphics[width=4.8cm]{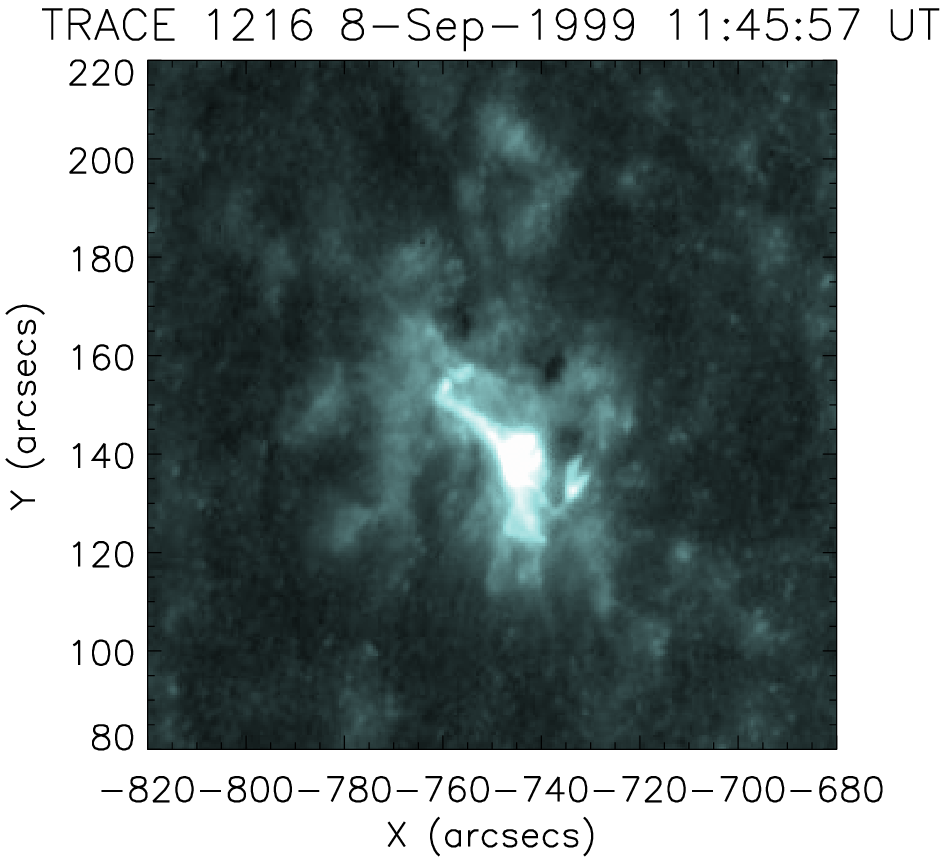}
 }
\vspace{-0.25cm}
\hbox{
 \includegraphics[width=4.8cm]{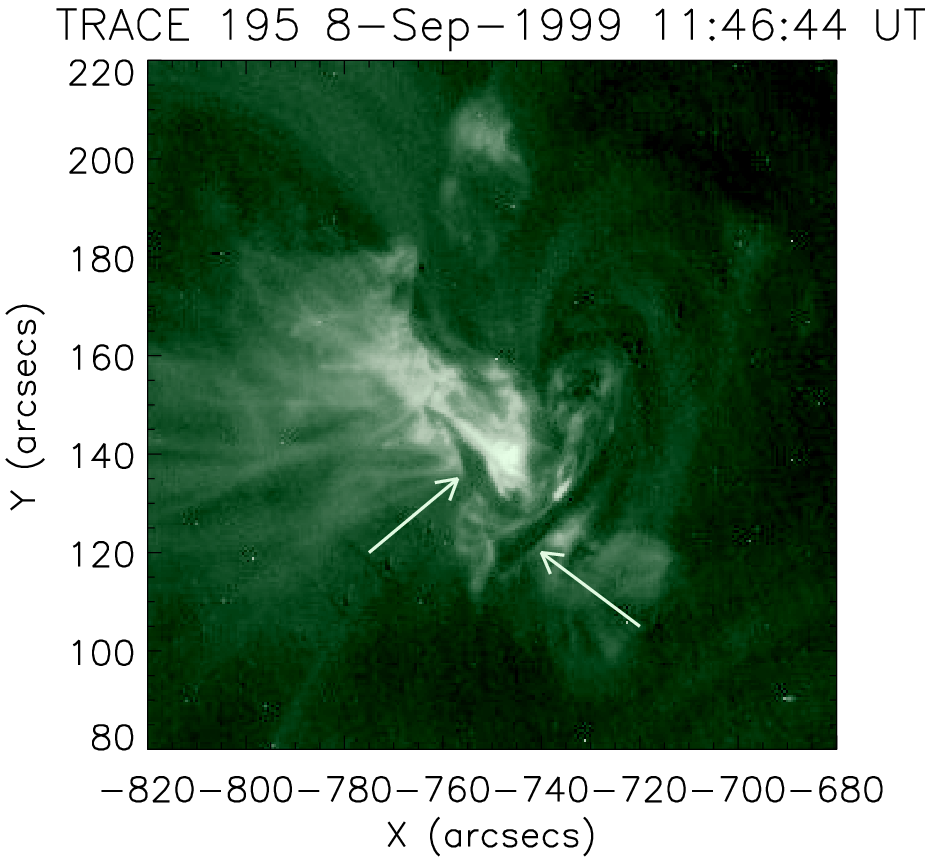}
 \hspace{-0.5cm}
 \includegraphics[width=4.8cm]{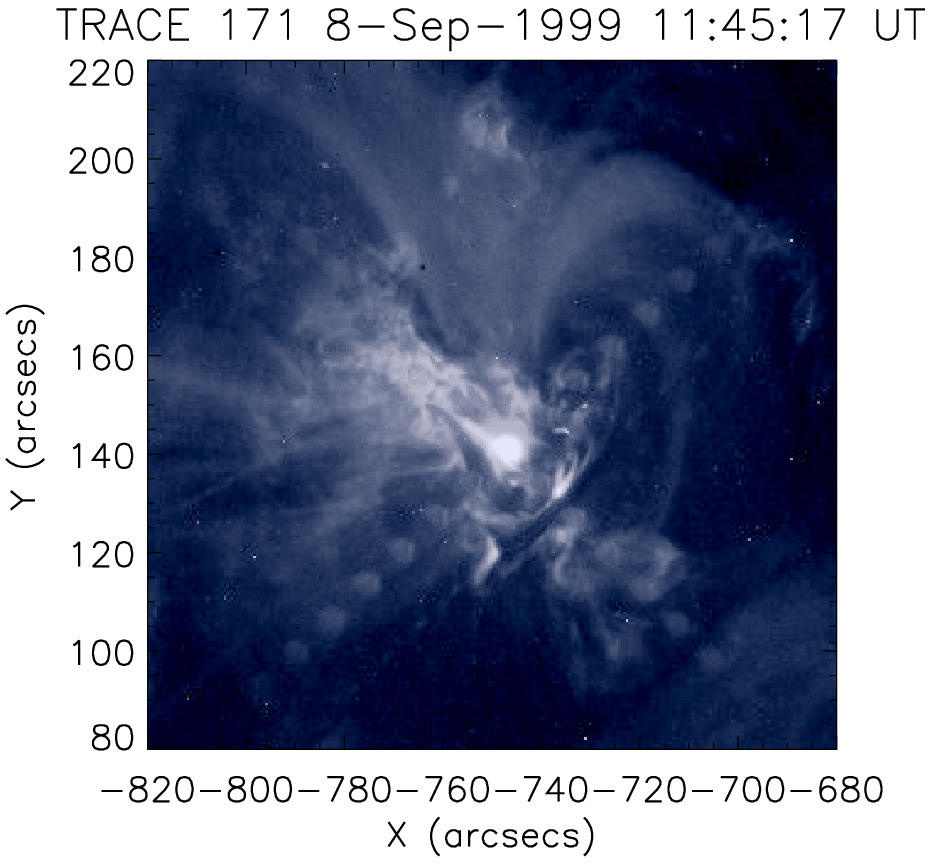}
}
 \caption{Upper panels: TRACE WL image showing the photospheric configuration of NOAA 8690 on Sep 8, 1999 and the TRACE 1216~\AA\ image from the same time. Lower panels: TRACE EUV images showing the two small filaments present in the active region just before the flare, indicated by the arrows on the lower left panel.}
     \label{fig_wl}
\end{figure}

Active region NOAA 8690 appeared on 05-September-1999 as a small beta region on the north east of the limb of the Sun with arch type filaments and bright H$\alpha$ plage, and small sunspots. Fig. \ref{fig_wl}  (upper left panel) shows its configuration before the M1.4 flare, which is a sunspot group characterised by a $\beta \gamma \delta$ magnetic configuration (from the  BBSO Solar Activity Report). TRACE images in the 171~\AA\ and 195~\AA\ wavelengths from just before the flare show two small filaments (Fig. 5, lower panels). The more \begin{bf}eastern\end{bf} of these filaments is visible also in emission in the \lya\ channel, unlike the eastern filament, which is not clear either in emission or absorption. 

\begin{figure*}
\centering
  \includegraphics[width=16.5cm]{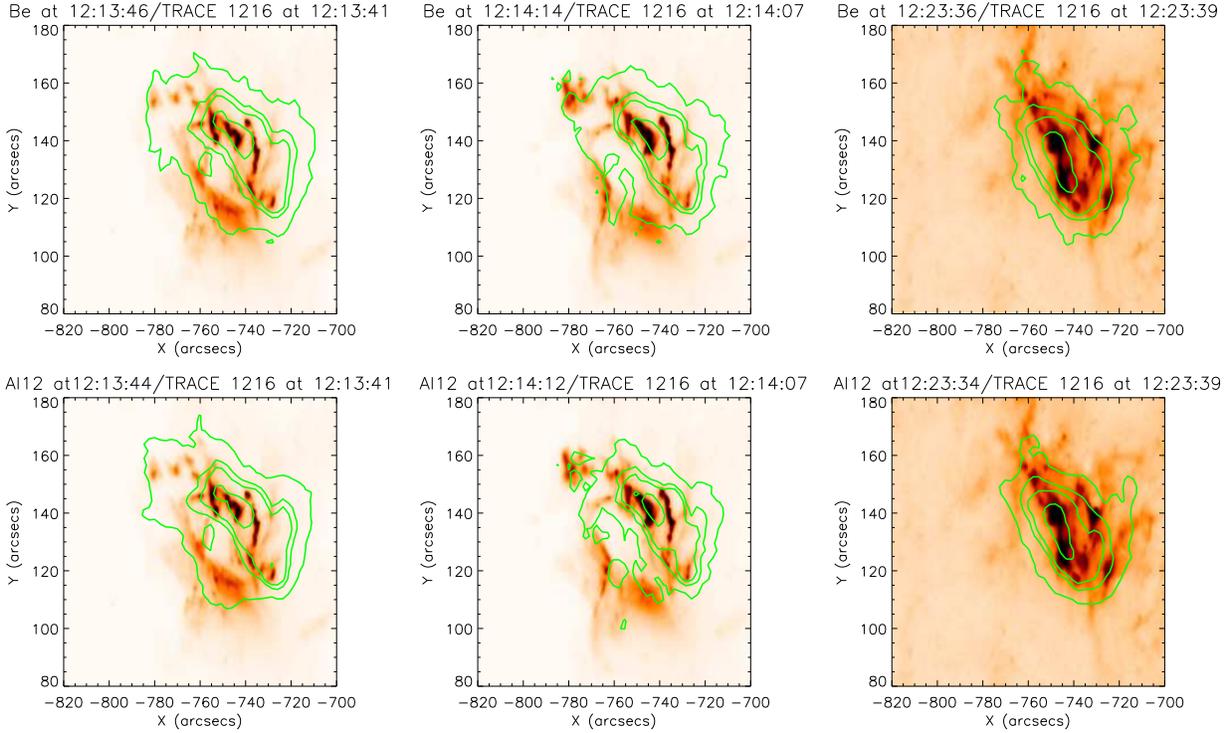}
     \caption{Temporal evolution of the Yohkoh/SXT source (green contours), superposed on the nearest TRACE 1216~\AA\  imges. The upper panels show Be filter images, and the lower panels show Al 12 filter images. The contour levels are 1\%, 5\%, 10\% and 50\% of the peak counts in each SXT image.}
     \label{fig4}
\end{figure*}

The event starts around 12:05~UT with bright ejecta seen in TRACE 1216~\AA. A pair of TRACE ribbons start brightening approximately 12:13:07, with a maximum at 12:14:23. The ribbons visible in Figure~\ref{fig4} spread apart, and the ejecta remain bright until around 12:16:30 UT, after which they become faint and diffuse, and difficult to distinguish from the background plage (see difference images in Figure~\ref{fig:1216ejecta}). Between the brightest ejecta, seen in the left and middle columns of  Figure~\ref{fig4}, and the footpoints there appears to be a void with a well-defined upper boundary, implying the ejection of a discrete structure rather than a continuous outflow. The cloud of ejecta, most clearly visible to the south and east of the ribbons, does not appear like expanding loops or a confined jet, or a coherent dense filament. It has the appearance of a spreading semi-circular sheet of material with embedded `blobs'. This type of morphology is also seen in around 25\% of flares observed in soft X-rays, and was classified as a spray by  \cite{2004JKAS...37..171K}. In the final phase of the flare, TRACE 1216~\AA\ images show two obvious ribbons slowly separating.

Fig.~\ref{fig4} shows the overlay of soft X-ray emission on 1216~\AA\ images.  The first two images in each row are from the impulsive phase,  at 12:13:42 UT and 12:14:02 UT, and the third is from the gradual phase. The bright emission seen in the TRACE 1216~\AA\ channel is spatially well correlated with the strongest emission seen in the SXT images. As the main emission in TRACE appears to come from footpoints, this is suggestive of early phase soft X-ray footpoints \citep[as observed previously by][]{1994ApJ...422L..25H}. Also apparent from these images is a faint SXR emission to the south-east of the strong footpoints, and co-spatial with the brightest part of the ejecta. This might imply that both hot and cool plasma are present in the ejecta. Later in the event the brightest SXR emission is co-spatial with the eastern ribbon. However, its position is also consistent with what would be expected of emission from the top of an arcade of hot flare loops joining the two sets of footpoints, as is usually found in the later phase of flares.

\begin{figure*}
\centering
  \includegraphics[width=16.5cm,height=10cm]{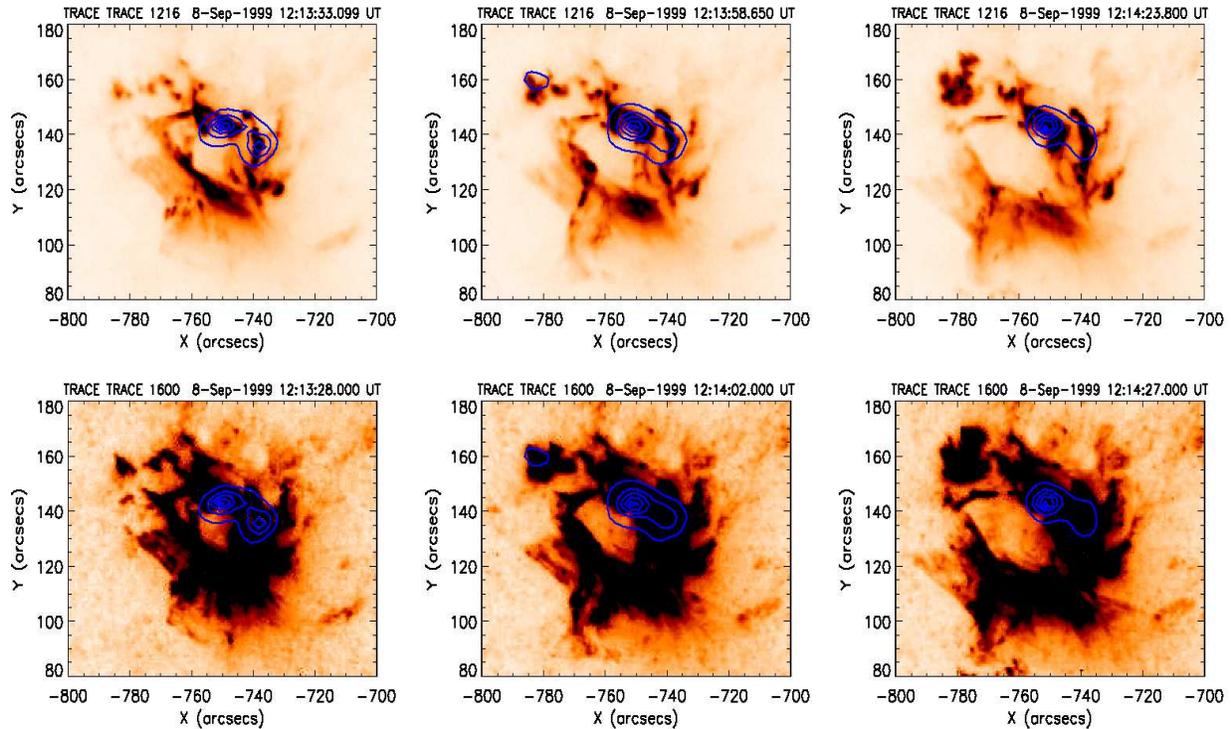}
     \caption{The 14-93 keV HXR source (blue contours), superposed on the nearest TRACE 1216~\AA\ 
(upper) and 1600~\AA\ (lower) images.}
     \label{fig5}
\end{figure*}

\begin{figure*}
\centering
  \includegraphics[width=16.5cm]{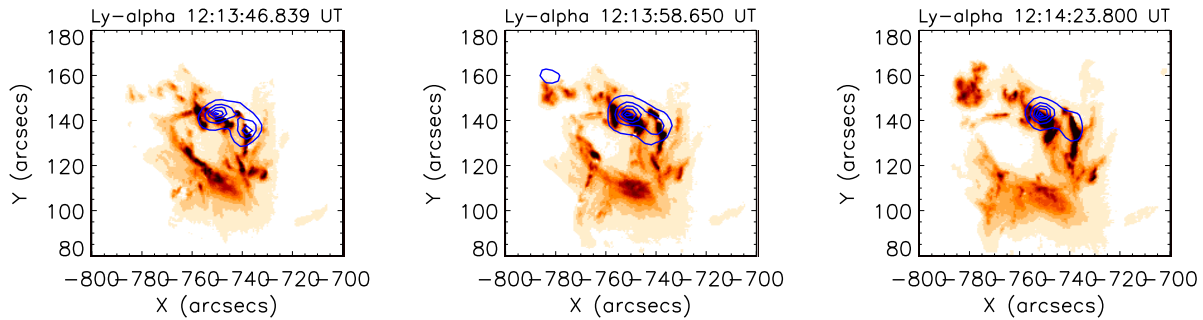}
     \caption{Temporal evolution of the \textbf{14}-93 keV source (blue contours), superposed on the nearest L$\alpha$
images corrected for ultraviolet contamination around 1600~\AA\, using equation (\ref{eq1}).}
     \label{fig8}
\end{figure*}

 Fig. \ref{fig5} shows the hard X-ray flare sources overlaid on 1216~\AA\ (upper panel) and 1600~\AA\ (lower panel) TRACE emission. The HXR images are produced using the MEM algorithm in the \emph{Yohkoh}/HXT software, integrating HXR counts from \begin{bf}14\end{bf}-93 keV. The HXR and TRACE 1216~\AA\ footpoints again coincide well.
Fig. \ref{fig8} also shows an overlay of HXR emission on the nearest \lya\ images, this time corrected for the UV contamination using Eq. (\ref{eq1}). The use of the corrected images allows us to better visualise the correspondence between HXR source and the flare ribbons seen in \lya.

\begin{figure}
\centering
\hbox{
  \includegraphics[width=4.8cm]{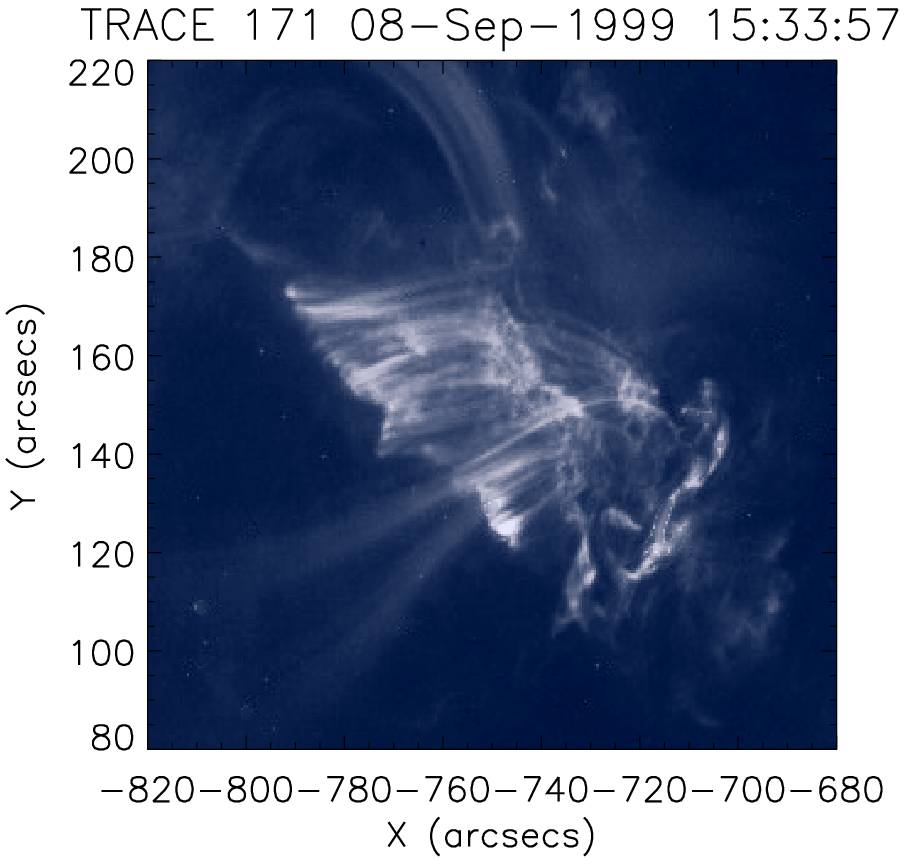}
  \hspace{-0.5cm}
  \includegraphics[width=4.8cm]{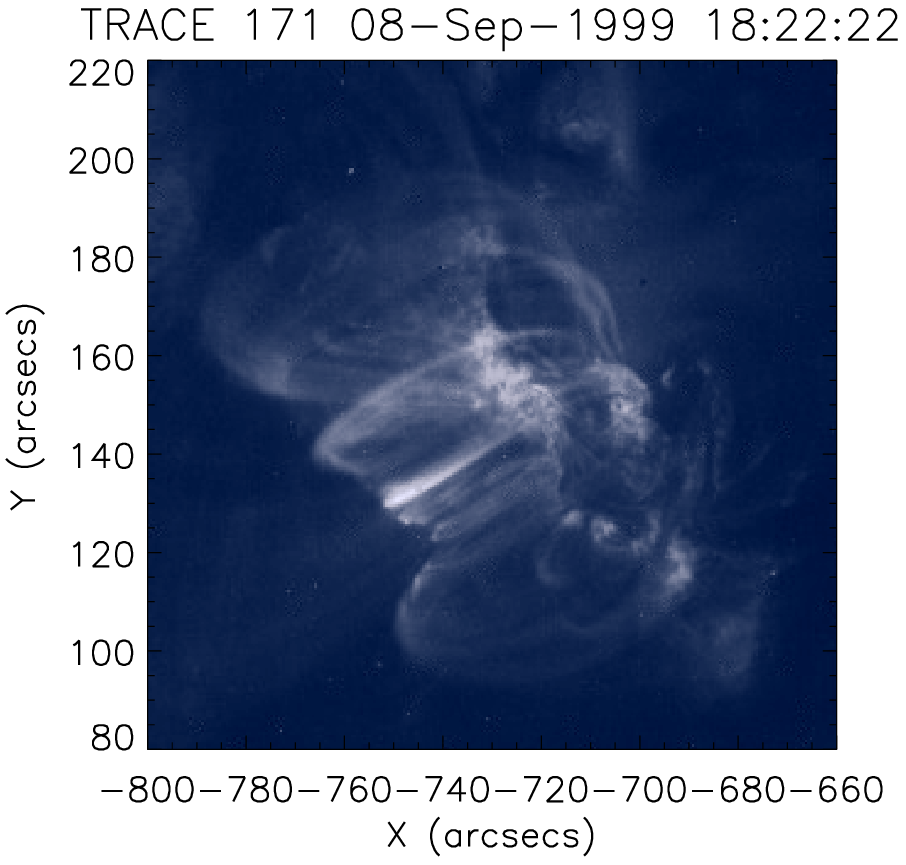}
 }
      \caption{171~\AA\ TRACE images from later on in the event.}
     \label{fig_171}
\end{figure}

Fig. \ref{fig_171} shows the evolution of the flare at 171~\AA. A few hours after the flare, a post-flare arcade is visible at coronal temperatures. As one would expect, the post-flare loops bridge the ribbons that are observed at 1216 and 1600~\AA, expanding into and beyond the `void' region that appeared below the UV ejecta, even taking the same shape as this void - compare loops in Fig. \ref{fig_171} left panel with void in \ref{fig5} top right.

\section{Flare Energetics from TRACE 1216~\AA\ and 1600~\AA\ observations}\label{sect:uvenergetics}

We use the corrected and normalised \lya\ images (i.e. with the UV contamination removed) to infer the total \lya\ power in the flare footpoints.   Using IDL SolarSoft routine \emph{trace\_uv\_resp} we can convert the corrected \lya\ DN s$^{-1}$ into an intensity measured in photons cm$^{-2}$ s$^{-1}$ sr$^{-1}$. We assume that correction removes the 1600~\AA\ component, leaving only the peak around 1216~\AA, where the response function gives us 1~DN s$^{-1}$ pixel$^{-1}$$=1.25\times 10^{14}$~photons cm$^{-2}$ s$^{-1}$ sr$^{-1}$. Using a characteristic photon energy corresponding to a \lya\ photon with  $h\nu=1.635\times 10^{-11}$~erg we can then convert from DN s$^{-1}$ pixel$^{-1}$ to erg cm$^{-2}$ s$^{-1}$ sr$^{-1}$.

As the intensity measured in the TRACE image also includes a contribution from the background, it is advisable 
to repeat the previously mentioned steps for a quiet sun region, in order to measure the background and subtract the 
intensity of the background from the intensity of the footpoints. This also gives an idea of how close the results are to the previous spectrally-resolved measurements of \citet{1981A&A...103..160L}. Results are presented in Table~\ref{table_photometry}. 

\begin{table}
\centering
\caption{Value of the corrected \lya intensity measured in a quiet Sun region and over a region in the flare footpoints.}
\begin{tabular}{c c c}
\hline
 & Quiet Sun region & Footpoint region \\
\hline
   counts $\rm{s^{-1}}$ & 7.35$\times 10^5$ & 3.66$\times 10^6$ \\
   counts s$^{-1}$ px$^{-1}$ & 68 & $5.64 \times 10^3$ \\
   photons s$^{-1}$ cm$^{-2}$ sr$^{-1}$ & 1.16$\times 10^{16}$ & 8.26$\times 10^{17}$ \\
   erg s$^{-1}$ cm$^{-2}$ sr$^{-1}$ & $1.63\times10^5$ & 1.35$\times 10^7$ \\
\hline
\end{tabular}
\label{table_photometry}
\end{table}

Using this empirical correction method, we determine the intensity of the quiet Sun at \lya\ to be $1.63 \times 10^5$ erg~cm$^{-2}$ s$^{-1}$ sr$^{-1}$.  In spectroscopic observations, \citet{1981A&A...103..160L} found the intensity of the quiet Sun at \lya\ to be $6.55 \times 10^4$ erg~cm$^{-2}$ s$^{-1}$ sr$^{-1}$ over a 2\AA\ passband, so our quiet Sun measurements are overestimating the intensity by a factor of 2.5. The reason for this is not clear, since in the quiet Sun the Kim et al. correction should be acceptable. It may be due to a changing calibration of the TRACE 1600~\AA\ and 1216~\AA\ channels due to degradation of the lumogen with time, or an incomplete removal of the continuum. It is likely that the flare measurements are also overestimates by about the same factor. The flare footpoint intensity in this event (GOES class M1.4) was $1.35 \times 10^7$~erg~cm$^{-2}$ s$^{-1}$ sr$^{-1}$. If this is also corrected by the factor 2.5, the resulting value of $5.4 \times 10^6$~erg cm$^{-2}$ s$^{-1}$ sr$^{-1}$ compares reasonably well with the value of 
$2.1 \times 10^6$~erg cm$^{-2}$ s$^{-1}$ sr$^{-1}$ measured by \cite{1984SoPh...90...63L} for an M1.0 event (which was a factor 1.4 less powerful than the 08-Sep-99 flare in terms of its soft X-ray flux). However, the discrepancy does suggest that the \cite{2006A&A...456..747K} correction does not sufficiently correct for enhanced C IV emission under flare conditions.

Assuming isotropic emission, the \lya\ power per unit area in the footpoint regions is $1.7\times 10^8$~erg s$^{-1}$ cm$^{-2}$ (or $6.8\times 10^7$~erg s$^{-1}$ cm$^{-2}$ applying the correction factor of 2.5).  The area of a TRACE pixel is about (325 km)$^2$, so that the emission is 1.8$\times 10^{23}$~erg s$^{-1}$ pixel$^{-1}$, or around $10^{26}$~erg s$^{-1}$ integrated over the entire footpoint region. As will be shown in Section~\ref{sect:hxrenergetics}, this is consistent with the electron beam power calculated from the {\it Yohkoh}/HXT observations assuming a collisional thick target.

The VAL3C model of \cite{1981ApJS...45..635V} and the F1 and F2 models of \cite{1980ApJ...242..336M} give a flare to quiet Sun ratio for \lya\ at Sun centre of about 50 and 350 respectively. Our values given in Table~\ref{table_photometry} seem to be more consistent with the F1 model which describes faint to moderate flares. \cite{1980ApJ...242..336M} note that in their F1 model, the \lya\ line is the most important contributor to the radiative losses at the top of the flare chromosphere. This underlines the need for more detailed modelling \citep[such as done by][]{2005ApJ...630..573A}, and moreover, for high spatial and spectral resolution observations. We plan to investigate more closely the TRACE \lya\ emission in our observations with respect to semi-empirical models of the flare atmosphere in a separate study.

\section{Flare Hard X-ray energetics}\label{sect:hxrenergetics} 

\emph{Yohkoh}/HXT observed this flare only for the first minute of the impulsive phase. However this provides enough data to calculate the electron beam power under the assumption of a chromospheric collisional thick target.  The four 
HXT energy channels L, M1, M2 and H showed strong impulsive bursts. We fit the photon spectrum with
\begin{equation}
\centering
I(\epsilon)=A\epsilon^{-\gamma}\ \mathrm{photons~cm^{-2}~s^{-1}~keV^{-1}},
\label{eq3}
\end{equation}
where $I$ is the photon flux at the energy $\epsilon$ in keV, $\gamma$ is the power law index and $A$ is a normalisation constant. It is then possible to deduce the instantaneous electron energy flux at injection, above some `cut-off' energy $E_c$, assuming that the power-law part of  the spectrum is generated by electron-proton bremsstrahlung in a collisional thick target process. Assuming the Bethe-Heitler cross section, the following can be used \citep{2007ApJ...656.1187F}:

\begin{equation}
\centering
P(E \geq E_c)=5.3\times 10^{24} \gamma^2(\gamma-1)B \big( \gamma-\frac{1}{2},\frac{3}{2} \big) AE_c^{1-\gamma},
\label{eq2}
\end{equation}
where $B$ is the beta function and $A$ is the normalization constant of the photon spectrum $I(\epsilon)$.

\begin{figure}
\centering
  \includegraphics[width=9.1cm]{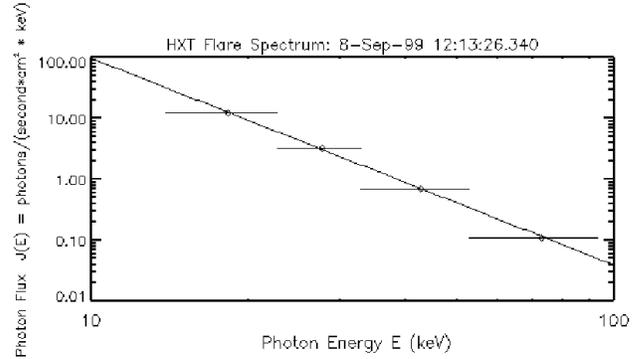}
     \caption{Hard X-ray spectra derived from the channel M2 of \emph{Yohkoh}/HXT data at the maximum of the flare.}
     \label{fig9}
\end{figure}

In Fig.~\ref{fig9}, we show the X-ray photon spectrum derived from the \emph{Yohkoh}/HXT data at the maximum of the hard 
X-ray flare. The spectrum is well fitted by a power law with index 3.40 which indicates a very 
hard spectrum. This result suggests that the hard X-ray emission above 14 keV was produced by nonthermal electrons. We tried fitting both a power law plus thermal spectrum and a single thermal spectrum, but the single power-law provided a significantly better fit.

The total energy in the electron spectrum depends on the value of $E_c$ which cannot be determined from this spectrum.   Instead we use three values of $E_c$ and calculate the instantaneous beam power carried by electrons above this energy. The spectrum implies that it might be reasonable to assume a low energy cutoff as low as 15~keV, though values of 20-25~keV are more typically assumed. In Table~\ref{table2}, we present the parameters derived from the single power law fit.

\begin{table*}
\centering
\caption{Parameters derived from \emph{Yohkoh}/HXT data.}
\begin{tabular}{c c c c c}
\hline
power law index & normalization constant & $P \geq E_c=15$ & $P \geq E_c=20$ & $P \geq E_c=25$ \\
$\gamma$ & $A$ & ($10^{27}$~erg s$^{-1}$) & (10$^{27}$~erg s$^{-1}$) & ($10^{27}$~erg s$^{-1}$)\\
\hline
3.399 $\pm$ 0.008 & $2.43 \pm 0.08 \times10^5$ & 8.61 $\pm$ 0.28 & 4.31 $\pm$ 0.14 & 2.53 $\pm$ 0.08 \\
\hline
\end{tabular}
\label{table2}
\end{table*}

The values in Table~\ref{table2} are reasonable for a small M-class flare such as this, and are also adequate to power the \lya\ losses calculated in Section~\ref{sect:uvenergetics}. 

\section{Large-scale dynamics -- the filament ejection and chromospheric disturbance}\label{sect:dynamics}

\begin{figure*}[!ht]
\centering
\begin{vbox}{
\includegraphics[width=18cm]{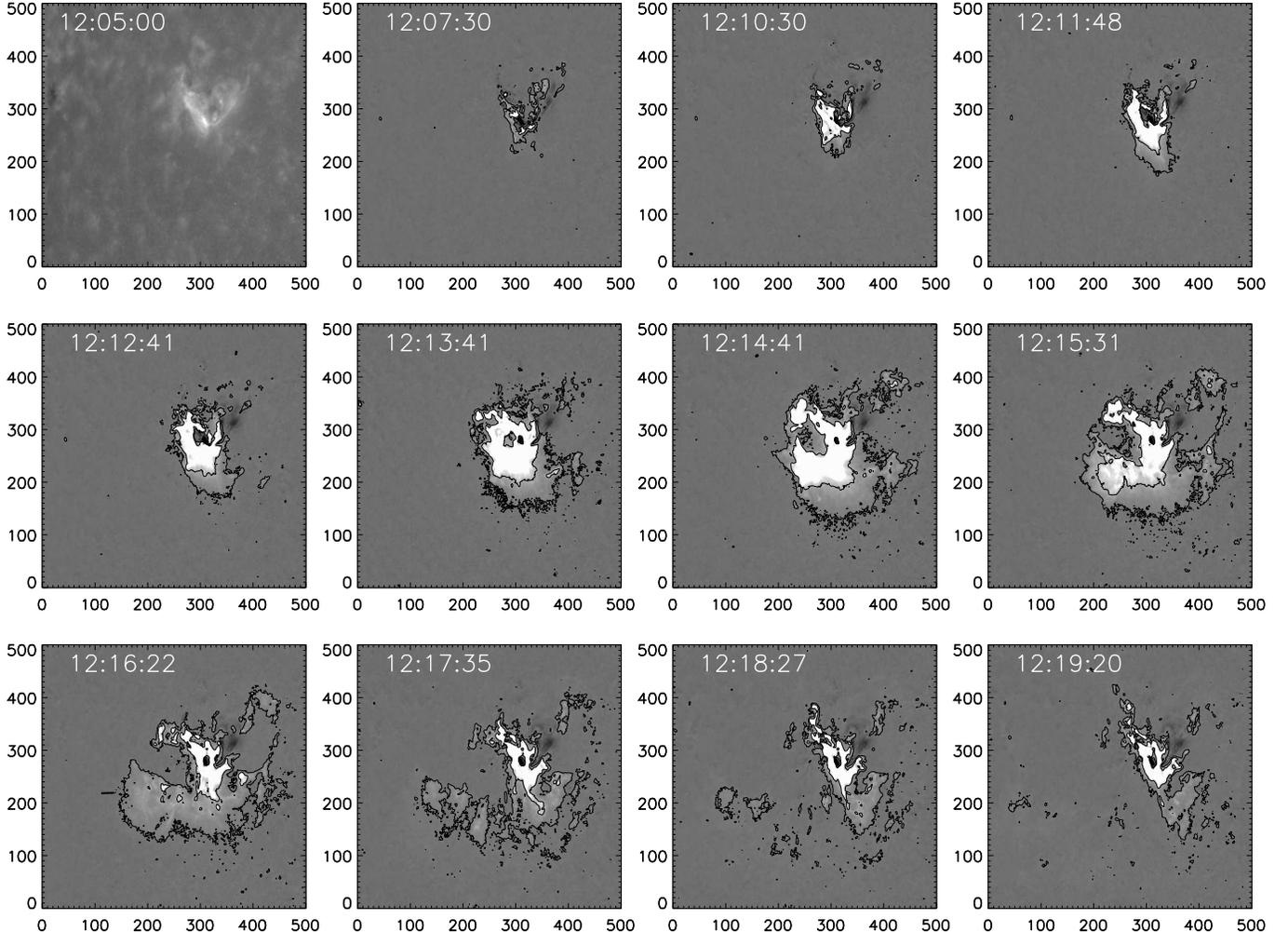}
}
\end{vbox}
\caption{1216~\AA\ base difference images  showing evolution of the erupting filament. The range shown from white to black is $\pm 200$~DN/s. The top left image is the base image.}
\label{fig:1216ejecta}
\end{figure*}

The TRACE UV observations also let us examine the evolution of the extended flare region at very high spatial and temporal resolution. A series of base difference images from the 1216~\AA\ channel is shown in Figure~\ref{fig:1216ejecta} (uncorrected for UV contamination). The time of the base image is 12:05:00~UT, the time at which the 1216~\AA\ series began, and we have selected only a subset of the images to show the overall evolution. The flat-fielded and normalised TRACE images are first cross-correlated to correct for drift and solar rotation, and then the base image is subtracted from following images. Therefore, parts of the difference image appearing bright are enhanced emission compared to the base image. To the north west of the main flare brightening, at image pixel location $\sim$ (360, 300) there is a dark region corresponding to the disappearance of some loops just visible in the first image. This could be due to the loops being ejected from the field of view within a minute or so (as no traveling bright feature is seen),  losing mass in some other way, or being heated out of the 1216~\AA\ passband. The ejecta appear as a ragged cloud expanding to the south east. This starts off bright and then fades in each subsequent image, presumably as the cloud expands. Around 12:14:41~UT and for a couple of minutes thereafter two patches at pixel location $\sim$ (400, 220) and (400, 400) become bright.

In the three images following 12:15:31 UT, the brightenings form a rough, diffuse semicircle, reminiscent of an `EIT wave' but very much fainter. The cloud expanding to the south east is structured on a scale of $\sim$ 20", with brighter `knots' within a more diffuse background. At 12:17:35, three or possibly four bright `knots' of emission are apparent in the cloud, moving to the south east. These are resolved by TRACE, and have a dimension at that time of approximately 20". The knots themselves also grow fainter and expand.  A rough estimate of the projected speed of the east-most knot is $300\pm50~\rm{km~s^{-1}}$.  Eventually the cloud of emission is no longer detectable above the background. We interpret the expanding brightenings to the south-east as moving, emitting gas. However the brightenings to the west are more static (e.g. around pixel position (400,420)), and may include some component of network emission which has been enhanced by the passage of the disturbance. 
Contour levels of [60, 200] DN/s in Figure~\ref{fig:1216ejecta} allow us, as previously, to estimate the surface intensity. A value of 60~DN/s corresponds to a surface intensity of approximately $1.2 \times 10^6$~erg cm$^{-2}$ s$^{-1}$.

The emission at \lya\ wavelengths suggests cool plasma in the ejecta, but as noted in Section 3, the ejection to the south east is also faintly visible in soft X-rays at around 12:13 - 12:14 UT. Unfortunately the SXR spatial resolution does not allow us to see in detail the spatial distribution of the 1216~\AA~ and SXR emitting plasma, so we cannot say whether they are co-spatial or are there hotter and cooler clumps. It is possible that the ejecta are primarily hot, with a small fraction of neutral H atoms efficiently scattering the intense radiation around 1216~\AA\ from the flare site. 
 
\section{Discussion and Conclusions}\label{sect:disc}
This paper presents the first examination of a flare in TRACE 1216~\AA~ at high spatial and temporal \begin{bf}resolution\end{bf}. We applied the empirical correction of \citet{1999SoPh..190..351H} in an attempt to separate the \lya\ radiation from the UV contamination in the 1216~\AA~channel, and in doing so have shown that the majority of the \lya\ emission originates from the flare footpoints, which are compact and co-spatial with the HXR sources. Erupting material is also visible in the corrected \lya\ images, as well as in the 1216~\AA~channel. The corrected \lya\ peaks at approximately the same time as the HXR (within a minute or so) but unfortunately we do not have sufficient time coverage in HXT or time resolution in the TRACE images to say more than this. From a knowledge of the TRACE response function we have calculated the power radiated in the \lya\ footpoints, and shown that this can be provided by the electrons which produce the observed HXR radiation, under the assumption of a collisional thick target chromosphere. 

The event was associated with the eruption of a small filament that was visible in the active region before the flare. During the eruption, the material emitted in soft X-rays as well as at the cooler TRACE wavelengths. The filament before the flare was presumably at temperatures of $\sim 10,000$ K, so the SXR emission implies that the filament heated as it erupted. The resolution of the SXT images is not sufficient to say whether the SXR emission is uniform or clumpy. In TRACE, the erupting material looked sheet-like with some embedded brighter knots which spread and fade. The erupting material could be mixed blobs of hot and cool material, or conceivably mostly heated filament material but still with sufficient neutral hydrogen to scatter UV flare radiation from the chromosphere. This might be the case if the hydrogen in the erupting filament does not have time to reach ionisation equilibrium.

In summary, this study has confirmed much of what was previously known about flares in \lya, including the typical \lya\ power radiated, and also demonstrated its close spatial and temporal association with the primary sites of energy release at the chromosphere. One of the most intriguing part of the observation is the filament eruption, which is already well underway before the flare footpoints brighten substantially, and which shows a morphology perhaps quite unlike what one would expect. It is neither rope-like - such as one might expect for the `core' of a classical 3-part CME, or bubble-like, such as one might expect for the `front' of a classical 3-part CME. In general, the flare events which are triggered by filament eruptions can be explained by the standard flare model CSHKP 
\citep{1964psf..conf..451C, 1966Natur.211..695S, 1974SoPh...34..323H, 1976SoPh...50...85K}.

The fact that the 1216~\AA~ channel and the corrected \lya\ emission both show the flare footpoints and the early stages of the onset of the eruption make this a promising channel for future observation of flares and eruptive events. Estimated \lya\ count rates are high from a small M flare such as this, which should enable high cadence imaging for similar or smaller events with future instruments.
As well as its intrinsic interest, we have been in fact motivated to perform this analysis by forthcoming instrumentation. The Extreme Ultraviolet Imaging Telescope \citep[see http://www.sidc.be/publications/docs/EUI-AthensPaper-20061120.pdf]{Hochedez} which has been selected for ESA's Solar Orbiter mission is intended to carry out \lya\ imaging, as will the LYOT imager on the proposed France-China SMESE satellite \citep{2007AdSpR..40.1787V} currently under study. These \lya\ instruments will provide an imaging overview of the flare evolution, both the flare ribbons/footpoints and the initial phases of the filament ejection, and will be particularly valuable scientifically when carried out in conjunction with hard X-ray imaging.

\begin{acknowledgements}
This work was supported by the European Commission through the SOLAIRE Network (MRTN-CT-2006-035484) and by STFC 
rolling grant STFC/F002941/1.
We would like to thank the referee for comments which helped to improve this paper.
\end{acknowledgements}

\bibliographystyle{aa}
\bibliography{lyman}

\end{document}